\documentclass[aps,prd,twocolumn,groupedaddress,eqsecnum,amssymb,showpacs,
nofootinbib]{revtex4}
\usepackage{graphicx}
\usepackage{mathptm}
\usepackage{bm}
\usepackage{times}
\usepackage{epsfig}

\begin{document}


\title{A Fly in the SOUP}

\author{R.~Holman}
\email[]{rh4a@andrew.cmu.edu}
\affiliation{Department of Physics, Carnegie Mellon University, Pittsburgh PA 15213, USA}
\author{L.~Mersini-Houghton}
\email[]{mersini@physics.unc.edu}
\affiliation{Department of Physics and Astrononmy, UNC-Chapel Hill, NC, 27599-3255, USA}

\date{\today}

\begin{abstract}
We investigate the Selection of Original Universe Proposal (SOUP) of Tye {\it et al} and show that as it stands, this proposal is flawed. The corrections to the Euclidean gravity action that were to select a Universe with a sufficiently large value of the cosmological constant $\Lambda$ to allow for an inflationary phase only serve to {\it renormalize} the cosmological constant so that $\Lambda\rightarrow \Lambda_{\rm eff}$. SOUP then predicts a wavefunction that is highly peaked around $\Lambda_{\rm eff}\rightarrow 0$, thereby reintroducing the issue of how to select initial conditions allowing for inflation in the early Universe.
\end{abstract}


\maketitle

\section{Introduction}
\label{sec:intro}

A theory of everything is not enough. Such a theory, be it string theory or something else, might allow us to understand the dynamical evolution of the Universe. However, without a theory of the {\em initial conditions} (IC), we will be severely restricted in the questions we can ask. While initial conditions for the Universe as a whole have been discussed extensively, especially in the context of quantum cosmology\cite{vilenkin,linde,hh}, the recently developed landscape picture of string vacua\cite{landscape} forces the issue to the fore. 

The unimaginably large number of string vacua found\cite{landscape} has been taken by some as a signal that anthropic arguments\cite{anthropic,nima} are the only ones that could be used to make predictions in string theory. If true, finding physical quantities that string theory might be able to predict becomes a hard, perhaps impossible task, at least until probability distributions peaked around universes like ours can be sensibly derived. 

On the other hand, not all physicists are comfortable enough with anthropic reasoning to give up on finding a more dynamical approach to vacuum selection in the landscape. There have been a number of attempts recently to do just this\cite{land1,land2,vafa,tye1,tye2}, mostly by trying to construct the relevant wave function of the Universe, or perhaps more appropriately, the wave function of the multiverse. The variables on which such a wave function should depend on would be those describing the landscape.
Presumably, if the wave function propagating on the landscape background contains information about the observable parameters which specify the Universe on large scales, we could use the probability distribution obtained from this wavefunction to make predictions for the values of these parameters\cite{land1,land2}. In Refs.\cite{land1,land2}, the scattering of the wavefuntion of the universe on the landscape background was treated as an $N$-body problem with solutions found over the whole multitude of landscape vacua. The probability distribution derived from such solutions was peaked around the universes with small energies.

The proposal put forth in Refs.\cite{tye1,tye2} considers the wavefunction to tunnel from a false to a true vacuum with the claim that the probability distribution is peaked around universes with reasonably large and finite $\Lambda$ when perturbations are taken into account. The Universe is supposed to appear from ``nothing'', that is,  a state with no classical notion of spacetime and then evolve into the Universe we see today. This sidesteps the issue of initial singularities present in the backward extrapolation of the classical cosmological spacetime.

To make this approach work, in terms of being able to identify which initial state is preferred by the wave function of the Universe, at least two conditions must be met. First, the wave function should be able to tell the difference between the various vacua. Second, in order to be able to compare the probabilities for different vacua, the wavefunction should be normalizable. If we look at the Hartle-Hawking (HH) ``no-boundary'' wave function\cite{hh}, we see that the latter condition does not occur. In particular, consider the HH wave-function corresponding to tunneling from ``nothing'' to a de Sitter space with cosmological constant $\Lambda$. The semi-classical approximation gives
\begin{equation}
\label{eq:HH}
\Psi_{HH} \sim \exp\left( -S_E\right) =\exp\left(\frac{3 \pi}{2  G_N \Lambda}\right).
\end{equation}
Here $S_E$ is the Euclidean action for the instanton that dominates the path integral with the relevant boundary conditions. We see that such a wave function indicates a preference for {\em low} cosmological constants, or equivalently, large horizons. We also see that by making $\Lambda$ smaller and smaller the Euclidean action can be made as negative as we want, which renders the HH wave function non-normalizable. This is a standard problem of Euclidean gravity; the conformal mode of the metric corresponds to a runaway direction in the superspace\cite{DeWitt} of Euclidean gravity. 

This calculation has something to teach us. As it stands, we have only included the cosmological constant as one of the physical parameters we would want to predict from the dynamics of the wave function. While important, we want more; we want the values of the other parameters that specify where we are in the landscape. Furthermore, the calculation would have us believe that the most probable Universe has $\Lambda=0$. The SNIa data do not support this, and in any case, this would also eliminate the possibility of an inflationary phase in the evolution of the Universe. The WMAP data are certainly consistent with inflation, and may in fact require an inflationary phase with an energy scale near the GUT/string scale. 

Here we investigate the proposal of Tye and collaborators\cite{tye1,tye2} (henceforth known as I, II, respectively) which offers a possible resolution to both of the points described in the previous paragraph. They call this proposal SOUP for ``Selection of the Original Universe Proposal''. Their proposal is based on going beyond the strict minisuperspace approach for computing solutions to the Wheeler-DeWitt (WDW) equation,  including matter fluctuations as well as metric perturbations, which tend to decohere the wave function. This has the effect of suppressing the tunneling amplitude in a way that depends on the field content of the theory, as well as on the parameters that specify which state the Universe will tunnel into. An example of this would be the KKLT\cite{kklt} and KKLMMT\cite{KKLMMT} scenarios in which the de Sitter state is generated through a combination of non-trivial fluxes and brane-antibrane configurations. If the wave function depends on the fluxes, then we would be able to use the wave function to predict which compactification will be preferred. 

The essence of the Tye {\it et al} idea is to argue that when tunneling to a de Sitter space, the decoherence effects on the wave function will in general prefer values of the cosmological constant that are {\em generic}, neither too large, nor too small, relative to the natural energy scales in the theory, presumably the Planck/string scale here. The vacuum energy $\Lambda$ is  the brane tension $\sigma$ obtained by stacking $N$ pairs of branes: $\Lambda \simeq 2N\sigma$. The HH wave function changes from that in Eq.(\ref{eq:HH}) to
\begin{equation}
\label{eq:modHH}
\Psi \sim \exp {\mathcal{F}},\ {\mathcal F} = -S_E-{\mathcal D},
\end{equation}
with $\mathcal{F}$ the effective action obtained after coarse-graining and where the decoherence correction term ${\mathcal D}$, which is real and positive, originates from the influence functional obtained by tracing out the metric and matter perturbation modes with wavelengths larger than the cuttoff size $\sqrt{\Lambda}$. In I the authors argue that ${\mathcal D}$ is proportional to the area of the boundary towards the end of tunneling so that  ${\mathcal D}$ is a {\it constant} depending on the cuttoff scale $\Lambda$ (since the size of boundary is given by the initial de Sitter horizon $\sqrt\Lambda$). For a $D$-dimensional inflationary universe with cosmological constant $\Lambda$ this procedure yields,
\begin{equation}
\label{eq:decohlam}
{\mathcal F} \simeq \frac{a}{\Lambda^{\frac{D-2}{2}}} - \frac{b}{\Lambda^{\frac{D-1}{2}}}.
\end{equation}
where $D$ denotes the dimensionality of space.
Minimizing this for generic values of $a,\ b$ gives generic values of $\Lambda$. In II, metric perturbations with wavelengths larger than the horizon are traced over, a procedure which yields a {\it radiation-like} correction term to the effective action of the form, ${\mathcal D} \simeq {\nu}\slash {\Lambda^2 a^4}$.

SOUP is an interesting proposal and it provides yet another example of non-anthropic selection of vacua from the multitude of vacua in the landscape. However, as it stands, we do not think it gives a correct determination of the cosmological constant as the argument above would suggest. We will argue this point in the rest of the paper and comment on an alternative criterion for vacuum selection based on gravitational instabilities of perturbations; we will pursue the derivation and consequences of this criterion in a second paper\cite{richlaura2}.

In the next section, we flesh out the part of the discussion of Tye {\it et al} that we take issue with. In Section \ref{sec:tyeprob} we show how the SOUP proposal, at least as currently envisaged,  does {\em not} select out a value for $\Lambda$ other than the one selected by the original HH wave function, {\it i.e.} $\Lambda\rightarrow 0$. We then turn to a discussion of our proposal  and conclude in Sec.\ref{sec:conclusions}.

\section{The Problems with SOUP}
\label{sec:tyeprob}
The claim in I is that the inclusion of the inhomogeneous modes of the metric and/or matter fields will change the $\Lambda$ dependence of the wave function of the Universe, at least in the semiclassical approximation, in such a way that a preferred {\em non-zero} value is selected. In essence, an effective potential for $\Lambda$ is generated that has a minimum away from $\Lambda=0$. The parameters of this potential depend on the de Sitter boundary size $\Lambda$ and the number of the degrees of freedom $b$ of the theory, so that one can correlate a given value of $\Lambda$ as picked out by minimizing the effective potential with values of the parameters of theory. As an example of this, Ref.\cite{huang} shows how the modified wave function can be used to argue that the number of e-folds in chaotic inflation models is most likely to be equal to the minimum number of e-folds ($\sim 60$) required to solve the horizon and flatness problems. 

The first problem we see with this proposal has to do with whether or not we can {\it distinguish} between the  original value of $\Lambda$ appearing in the action and the effective potential generated for it. Recall the prescription for constructing the HH wave function. We are to compute the following Euclidean path integral
\begin{equation}
\label{eq:PIHH}
\Psi_{\rm HH}\left[h_{ij}; \phi\right] = \int {\cal D}g_{\mu \nu}\ {\cal D}\Phi \ \exp\left(-S_E\right),
\end{equation}
where the geometries involved are {\em compact} four-geometries having metric $h_{ij}$ as their boundary and the fields are fixed to the value $\phi$ on this boundary. The HH wave function relevant to the creation of a de Sitter Universe is obtained by evaluating the path integral in the semiclassical approximation, where the action is saturated with a solution to the classical Euclidean equations of motion with the appropriate boundary conditions; this is the so-called de Sitter instanton 
\begin{equation}
\label{eq:deSinst}
a(\tau) = \frac{1}{H} \cos H\tau,
\end{equation}
where $\tau$ is Euclidean time and for the instanton, we have the restriction $|\tau|\leq \pi\slash 2 H$. 

In the minisuperspace approximation on the landscape background $\phi$, where we only keep the homogenous mode $a(t)$ we arrive at the standard result in Eq.(\ref{eq:HH}). Now suppose we go beyond this approximation by including the inhomogeneous modes $\left\{x_n\right\}$. As shown in Ref.\cite{hawkinghalliwell}, to first order the wavefunctional is $\Psi[a, \phi] =\Psi_{0}(a,\phi)\ \prod_{n} \psi_{n}(a,\phi,x_n) \simeq e^{-S_0 + \Sigma_n S_n x_n^{2} }$ with $S_0$ being the unperturbed action and $S_n=S_n\left(a, \phi\right)$ the correction terms arising from the perturbations. Tracing out the $\left\{x_n\right\}$ then yields  an effective action $\mathcal{A}_E$ and wavefunction $\Psi(a,\phi)\simeq e^{- \mathcal{A}_{E}/2 }$ and generates the reduced density matrix for the homogeneous modes $\rho_{\rm red}\left(a,a^{\prime}\right)$. This can then be used to compute the probability to tunnel from $a=0$ to $a=H^{-1}$. The effect of integrating the inhomogeneous modes out is to modify the Euclidean action from the given one $S_E$ to the so-called coarse grained effective action $A_E = \mathcal F$ \cite{rich}. Let us recall that in I, the effective action obtained after after tracing out modes with wavelength larger than the instanton boundary size $\sqrt{\Lambda}$, takes the form

\begin{equation}
\label{eq:decohlam1}
{\mathcal F} \simeq \frac{a}{\Lambda^{\frac{D-2}{2}}} - \frac{b}{\Lambda^{\frac{D-1}{2}}} = \frac{a}{\Lambda_{\rm eff}^{\frac{D-2}{2}}}.
\end{equation}
In II the tracing out of the tensor metric perturbations results in an effective action that contains a 'radiation-like correction:
\begin{equation}
\label{eq:radiation}
\mathcal{A}_E = \frac{1}{2}\int d\tau \left(-a\dot{a}^2 -a +\Lambda a^3 + \frac{\nu}{\Lambda^2 a} \right)
\end{equation}

The procedure would then be to look for solutions to the Euclidean equations of motion generated by varying the effective action $A_E$ with respect to minisuperspace variables $(a,\phi)$. At this point we need to be aware that $A_E$ actually depends on two scale factors $a,\ a^{\prime}$, since it gives rise to the propagator for the reduced density matrix. If the system has not decohered then variation  of $A_E$ with respect to $a,a'$ would yield two Friedman-like equations with different expansion rates $H (\Lambda) , H' (\Lambda_{\rm eff})$. Quantum entaglement would allow us to distinguish between $H,H'$ and therefore between the leading and the correction term in $A_E$. To what extent can we treat $A_E$ as only depending on one scale factor so that we can go through the same procedure as we would have in the de Sitter instanton case? This relates to the issue of whether the ``histories'' of the scale factors in question have small enough overlap so that each one can be treated classically, in the sense of not being interfered with quantum mechanically. To understand when this happens, we can make use of the results in Ref.\cite{halliwell,halliwell1} for the reduced density matrix $\rho_{\rm red}\left(a,a^{\prime}\right)$:
\begin{equation}
\label{eq:reddensmatrix}
\rho_{\rm red}\left(a,a^{\prime}\right)\sim \exp\left(-N\frac{\left(a+a^{\prime}\right)^2\left(a-a^{\prime}\right)}{4 a^2 {a^{\prime}}^2}\right)
\end{equation}
Here $N$ is the number of modes included in the environment that has been traced out. For large enough $N$, the reduced density matrix only has support for $a^{\prime}\sim a$ which is the sign that the quantum interference between these scale factors is small enough that we can use the classical equations of motion obtained from $A_E$ when we set $a=a^{\prime}$ in order to obtain the new form of the instanton relevant to the creation of a de Sitter Universe from ``nothing''. 

This brings us to the crux of our argument. The corrections to the original Euclidean action induced by tracing out degrees of freedom other than the scale factor will depend on $\Lambda$ in some non-trivial way. In I the correction terms do nothing more than to renormalize the cosmological constant $\Lambda$ to $\Lambda_{\rm eff}$ given by Eq.(\ref{eq:decohlam1}). The gravitational degree of freedom $a$ is the first one to become perfectly classical, $a=a'$. It is straightforward to prove this is the case by deriving the equations of motion for the Euclidean scale factor $a(\tau)$  
\begin{equation}
\label{eq:heff}
\dot{a}^2+1 = H_{\rm eff}^2 a^2+\cdots,
\end{equation}
where $\cdots$ indicate terms that might come from higher curvature operators induced by the tracing out procedure. Here $H_{\rm eff}= H_{\rm eff}\left(\Lambda\right)$ is nothing other than the effective potential for $\Lambda$! The net effect of tracing out the environment modes is to {\em reset} the cosmological constant to $\Lambda_{\rm eff} = 3 H_{\rm eff}^2$. We can understand this result as follows. There is only one scale factor $a$ and therefore only one expansion rate $H_{\rm eff}$ obtained by varying $A_E$ with respect to $a$. The quantity measured in this de Sitter universe can only be $\Lambda_{\rm eff}=3 H_{\rm eff}^2$. Unfortunately there is no way to distinguish between $\Lambda$ and $\Lambda_{\rm eff}$ or between the two constant contributions, (leading and correction term) in $A_E$, Eqs.(\ref{eq:radiation},\ref{eq:decohlam1}) by measuring $H_{\rm eff}$. The universe born out of the coarse-grained wavefunction given in terms of $\mathcal F$, Eq.(\ref{eq:decohlam1}), expands with a scale factor $a(t)$ given by Eq.(\ref{eq:heff}) and Hubble constant $\sqrt{\Lambda_{\rm eff}}$. Instead of determining a new value for the cosmological constant, we have just returned to our starting point; the entropy still scales as inverse of the renormalized cosmological constant, $\mathcal F = A_E \simeq (\Lambda_{\rm eff} )^{-1}$ and the improved HH wave function still prefers a Universe with zero cosmological constant.

In II the correction term has a different time dependence and evolution from the first term, Eq.(\ref{eq:radiation}), which might seem promising in terms of separating conributions from $\Lambda$ and the correction term in $H_{\rm eff}$. The correction terms originate by tracing out tensor metric perturbations with wavelengths longer than horizon size and contribute to the effective action by an amount $H^2 x_n^{2}$. As shown in \cite{hawkinghalliwell} these terms are bound to be very small since they contribute to the CMB fluctuations
\begin{equation}
\langle\delta T / T\rangle \simeq \langle x_n^{2}\rangle .
\label{cmb}
\end{equation}
On these grounds, their strength is at least five orders of magnitude less than the leading term $\Lambda^{-1}$ in the effective action. For this reason, while tracing out super-Hubble metric perturbation modes produces a radiation-like term in the expansion equation, this does {\em not} alleviate the problem of the entropy of inflation. This is still dominated by the leading term  $\Lambda^{-1}$.

There is another indication that the cosmological constant would not be fixed by what is essentially a statistical argument. One way to see that $\Lambda\rightarrow 0$ should be the preferred value when we use Euclidean quantum gravity is to note that the path integral for Euclidean gravity does not converge; the conformal factor gives rise to an unstable direction in the action and the action can be arbitrarily negative. This is true regardless of whether matter fields are present or not. Now suppose we trace out the fluctuation modes to generate the coarse grained effective action. To the extent that this action is local, we can expand it terms of powers of the curvature tensor of the background. The leading terms will be the Einstein action, which will again have the same problem of unboundedness from below. This contradicts the arguments in II to the effect that decoherence will allow bound the action, and we should expect a renormalization of the cosmological constant but nothing else.

\section{Conclusions}
\label{sec:conclusions}

The HH wave function argues for a Universe in which high scale inflation could not occur; from an entropic point of view, the entropy is given by $-S_E$ and use of the semiclassical approximation with the de Sitter instanton gives $S_E \simeq -1 / \Lambda$.

It has long been suggested that one possible resolution of this disrepancy would be to couple the system to other degrees of freedom, e.g. to the higher multipoles \cite{hawkinghalliwell,halliwell}. The effect of tracing out these environmental degrees of freedom may possibly address both of the above issues by adding corrections to the effective action of the system.

Tye et al.\cite{tye1,tye2} make use of this mechanism to argue that this procedure would select out potentially useful vacua for inflation from the string theory landscape. In I corrections to the effective action of the $(D-1)$ dimensional instantons, proportional to the area of their boundary were used as seen in Eqs(\ref{eq:decohlam},\ref{eq:decohlam1}).

In II radiative corrections were obtained by tracing out higher metric multipoles $\left\{x_n\right\}$ yielding corrections of the form $\mathcal{D} \simeq{\nu}\slash{\Lambda^2 a^4}$, with the hope that the ``friction-like" influence of the coupling to the environment of the higher multipoles, would suppress the tunneling rate in a vacuum dependent way and bound the Euclidean action from below.

Here we have argued that the perturbative correction in I  simply  renormalizes the cosmological constant given by Eq.(\ref{eq:decohlam}), $\mathcal{F}\simeq (\Lambda_{eff})^{-1}$, thereby reintroducing the original problems and implications of the HH-wavefunction. We also showed that the radiative corrections $\left\{x_n\right\}$ in II are subleading by at least five orders of magnitude compared to the zeroth order term in the action, $S_E \simeq 1/\Lambda$ since their strength is bounded by temperature and density perturbations, $\langle \delta T / T \rangle \simeq \langle x_n^2 \rangle $. Therefore they are not {\it sufficient} to suppress the tunneling rate. On top of this their contribution redshifts away as $a^{-4}$ compared to $\Lambda$. For this reason neither the entropy nor the boundedness issues can be resolved by tracing out metric perturbations.

The selection of the initial conditions, especially the fact that Universes that inflate must start out with unnaturally low entropy and the arrow of time are all intertwined into one deep problem.  Many proposals have been put forth\cite{entropy1,entropy2,anthropic,nima,lee, albrecht}, based either on a more conservative coarse-graining procedure or on more speculative conjectures like the N-bound, holography, causal patch physics, the complementarity principle and anthropic selection. However this puzzle remains one of the deepest mysteries in nature. The aforementioned proposals are succesful when applied to Black Holes but appear problematic when applied to the early Universe.

We take up this issue of initial condition selection in a separate paper\cite{richlaura2} and argue that the initial condition problem cannot be meaningfully addressed through thermostatistical arguments when gravitational degrees of freedom are involved. The implicit assumptions of ergodicity and thermal equilibrium in thermostatistics are valid for matter degrees of freedom, hence the successful application to the Black Hole entropy. But these assumptions are most likely incorrect when applied to gravitational degrees of freedom. In Ref.\cite{richlaura2} we propose to challenge the ergodicity and equilibrium assumptions and treat the problem of the initial conditions as a dynamical out-of equlibrium phenomenon for the combined sytem of (gravitational+matter) degrees of freedom.

\begin{acknowledgments}
R.H. was supported in part by DOE grant DE-FG03-91-ER40682.
\end{acknowledgments}

\end{document}